\documentclass{mem}
\usepackage{natbib}\usepackage{txfonts}\usepackage{balance}
\usepackage{graphicx}
\usepackage[a4paper]{}
\idline{75}{282}
\begin{document}
\def\teff{$T\rm_{eff }$}
\def\kms{$\mathrm {km s}^{-1}$}

\title{
Galactic nuclei formation via globular cluster merging
}

   \subtitle{}

\author{
R. \, Capuzzo-Dolcetta
\and P. \,Miocchi
          }

  \offprints{R. Capuzzo-Dolcetta}

\institute{``Sapienza'' Universit\'a di Roma, 
Dipartimento di Fisica, 
P.le A. Moro, 5, I-00185 Roma, Italy.
\email{miocchi@uniroma1.it}
}

\authorrunning{Capuzzo-Dolcetta \& Miocchi}

\titlerunning{Galactic nuclei formation}

\abstract{
Preliminary results are presented about a fully self-consistent $N$-body
simulation of a sample of four massive globular clusters in close interaction 
within the central region of a galaxy. The $N$-body representation (with 
$N=1.5\times 10^6$ particles in total) of both the clusters and the galaxy 
allows to include in a natural and self-consistent way
dynamical friction and tidal interactions. 
The results confirm the decay and merging of globulars as a viable scenario
for the formation/accretion of compact nuclear clusters.
Specifically: i) the frictional orbital decay is $\sim 2$ times faster than that
predicted by
the generalized Chandrasekhar formula; ii) the progenitor clusters merge in less
than $20$
galactic core-crossing times ($t_b$); iii) the NC configuration keeps
quasi-stable at least
within $\sim 70 t_b$.
}
\maketitle

\section{Introduction}
Nuclear clusters are common across the Hubble sequence \citep{math,scho}. 
In particular, the ACS Virgo Cluster Survey \citep{cote06} shows many compact 
nuclei at the photocenters of many of early-type galaxies ($>66\%$ for
M$_V<-15$). Their half-mass radii ($r_h$) are in the $2 \div 62$ pc range, with
average 
value $\langle r_h \rangle = 4.2$ pc, and scale with the nucleus luminosity as 
$r_h \propto L_n ^{0.5 \pm 0.003}$.
Brighter ($M_V < -20.5$) core-Sersic galaxies lack resolved stellar nuclei.
The mean of the frequency function for the nucleus-to-galaxy luminosity  ratio
in 
 nucleated galaxies, Log $\eta = - 2. 49 \pm 0.09$ is {\it indistinguishable} 
from that of the Super massive black hole--to-bulge mass ratio,
 Log $(M_\bullet /M_{gal}) = -2.61 \pm 0.07$, calculated in 23 early-type
galaxies with detected supermassive black holes (SBHs).It is thus argued that
resolved stellar nuclei are the low-mass counterparts of nuclei hosting SBHs
detected in the bright galaxies. 
If this view is correct, then one should think in terms of central massive
objects, 
either SBHs or compact stellar clusters (CSCs), that accompany the 
formation and/or early evolution of almost all early-type galaxies. Comparing
the nuclei to the nuclear clustersǒof late-type spiral galaxies reveals a close
match in terms of size, luminosity, and overall frequency. 
A formation mechanism that is rather insensitive to the detailed properties of
the host galaxy properties is required to explain this ubiquity and homogeneity.

Another observational relevant point is the growing evidence of presence of 
very massive ($>10^7$ M$_\odot$) YOUNG star clusters in Antennae, Magellanic 
Clouds, M33, M82, Fornax dSph  \citep{frit,degr,fusi}, 
as well as OLD  \citep{har} in M87 and Virgo ellipticals.

\citet{haretal} indicate how up to a 40\% of the total mass in 
 GCS of brightest cluster galaxies is contributed by massive 
(p.d.mass $> 1.5\times 10^6$ M$_\odot$, in good  agreement with recent 
theoretical results by \citet{kra}.
Putting together these observational data (presence of resolved stellar 
nuclei in galaxies and likely initial presence of massive star clusters, raise
two questions and 
possible answers. The questions are: i) how are these \lq stellar\rq  nuclei
formed?,
and ii) why very massive stellar clusters are no more observed in many galaxies?
The (possible) answers could be resumed in: resolved stellar nuclei are formed
via massive stellar clusters merging, after a substantial orbital decay toward
the galactic central region. In the following we describe some of our work done
recently to give substance to this explanation.

\section{Nuclei as remains of merged Globular Clusters?}

Many papers dealt with dynamical friction on massive objects orbiting 
galaxies. It seems well ascertained that sufficiently massive and compact
clusters may decay towards their parent galaxy central region in a time
short respect to the Hubble time. The less symmetries in the galactci potential,
the quicker the orbital decay, tha depends (obviously) also on the
initial orbital energy and angular momentum. An enhancement of the
classical frictional deceleration caused by background stars is due to
tidal braking torque caused by the galactic potential. As a consequence, 
it has been proved in quite general cases that after less than 1 Gyr 
many GCs are limited to move in the inner galactic region. 
Do they merge and form a Compact Star Cluster? The first positive answer to this
question was given by \citet{tos} who were the first to examine the modes of
cluster merging in the inner region of M 31.
Here, we summarize the results presented in \citet{cdm08b} of a fully 
self-consistent $N$-body simulation of the close interaction of a 
sample of four massive globular clusters (GCs) in the central region of a 
galaxy.
Both the clusters and the galaxy are represented by mutually interacting 
particles, thus including in a natural and self-consistent way dynamical
 friction and tidal interactions. 
This study represents a substantial improvement in the analysis of the
frictional decaying and merging of GCs in galactic nuclear regions, a 
scenario first tackled by semi-analitical approaches \citep{tos,cd93} and 
then pursued by $N$-body experiments \citep{oh00,cdm08a}. 
Clarifying the role of the above-mentioned dynamical effects
is important also to understand the formation and origin of
Nuclear Clusters (NCs) \citep[e.g.][]{oh00,bekki04}.
\section{Models and Results}
\begin{figure}[t!]
\resizebox{\hsize}{!}{\includegraphics{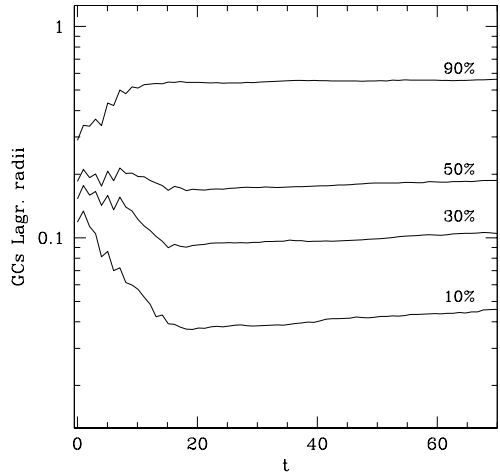}}
\caption{\footnotesize
Evolution of the Lagrangian radii of the four GCs as
a whole.}
\label{f1}
\end{figure}

Each GC is represented by 256,000 particles initially distributed according to a
King profile whose structural parameters are taken from the set of the
most compact clusters simulated in \citet{cdm08a}.
The GCs are initially located at rest within the galactic core (see
Fig.~\ref{f1}).
The galactic model is given by a spherical and isotropic Plummer phase-space
distribution
sampled with 512,000 particles.
The simulation is performed with our own parallel tree-code using individual and
variable
time-steps \citep{mio02}.

The simulation results can be re-scaled with any given set of galactic
structural parameters.
One possible choice for these parameters is the following: 
core radius $r_b=200$ pc;
core-crossing time $t_b=0.54$ Myr;
central density $\rho_{b0}=370$ M$_\odot$ pc$^{-3}$.

\begin{figure}[t!]
\includegraphics[width=6.5cm]{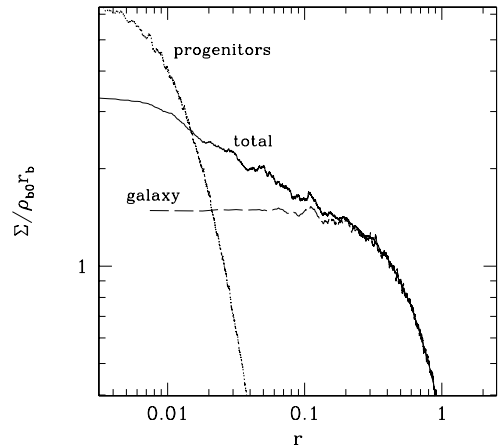}
\caption{\footnotesize
Projected surface density profiles of the 
final configuration of the whole system (galaxy plus NC) (solid line) 
and, for the sake of comparison, of the galaxy stellar component only
(long dashed line). The dotted line is the profile of the `summed'
four GC progenitors.}
\label{f3}
\end{figure}

The main results of the simulation can be summarized as follows:
i) the frictional orbital decay is $\sim 2$ times faster than that given by the
use of
the generalized Chandrasekhar formula; ii) the progenitor clusters (initially
located within
the galactic core) merge in less than $20$ galactic core-crossing time ($\sim
11$ Myr),
see Fig.~\ref{f1} and \citet{cdm08b}; iii) the NC configuration is quasi-stable
at
least within the simulated time ($\sim 70 t_b \sim 40$ Myr);
iv) the total surface density profile has the typical appearance of a nucleated
galaxy
central profile, see Fig. \ref{f3}; 
v) the global velocity dispersion profile \emph{decreases} towards the 
centre as found in the \citet{geha02} observations.
These results are described in more detail in \citet{cdm08b}.
%

\begin{figure}[t!]
\resizebox{\hsize}{!}{\includegraphics{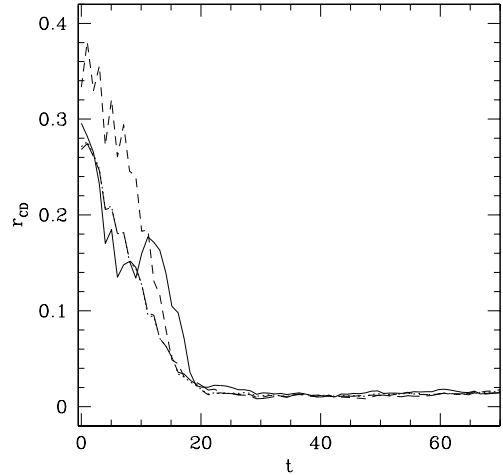}}
\caption{\footnotesize
Upper panel: time evolution of the distance of the
cluster CD from the galactic centre (rCD).}
\label{f2}
\end{figure}

\begin{figure}[t!]
\resizebox{\hsize}{!}{\includegraphics{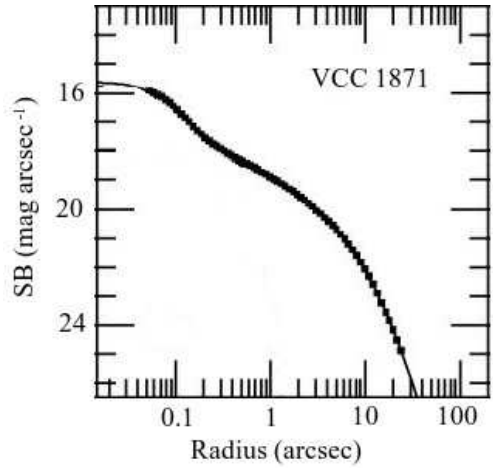}}
\caption{\footnotesize
Surface brightness (g-band) profile of the dwarf elliptical
galaxy VCC 1871 in the Virgo cluster (from \citep{cote06}).
}
\label{f4}
\end{figure}

\begin{figure}[t!]
\resizebox{\hsize}{!}{\includegraphics{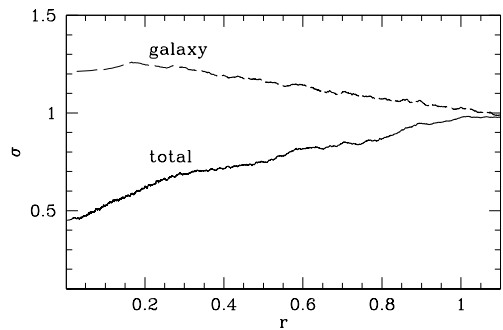}}
\caption{\footnotesize
Projected radial behaviour of the velocity dispersion
in the last system configuration. Line symbols are as in Fig. 4.}
\label{f5}
\end{figure}

\begin{acknowledgements}
The simulation was conducted at the CINECA supercomputing centre, under
the INAF-CINECA agreement (grant cne0in07).
\end{acknowledgements}

\bibliographystyle{aa}

\end{document}